\documentclass[a4paper]{jpconf}
\usepackage{amssymb,units,times,graphicx,bm,textcomp,amsmath,mcite}

\newcommand{\epsSlash}{/\!\!\!\epsilon}

\renewcommand{\e}{\textnormal{e}}

\renewcommand{\d}{\textnormal{d}}
\renewcommand{\i}{\textnormal{i}}

\begin{document}

\title{Novel aspects of radiation reaction in the classical and the quantum regime}

\author{Norman Neitz, Naveen Kumar, Felix Mackenroth, Karen Z. Hatsagortsyan, Christoph H. Keitel and Antonino Di Piazza}
\address{Max-Planck-Institut f\"ur Kernphysik, Saupfercheckweg 1, 69117 Heidelberg, Germany}

\ead{dipiazza@mpi-hd.mpg.de}

\begin{abstract}

This work is dedicated to the study of radiation reaction signatures in the framework of classical and quantum electrodynamics. Since there has been no distinct experimental validation of radiation reaction and its underlying equations so far and its impact is expected to be substantial for the construction of new experimental devices, e.g., quantum x-free electron lasers, a profound understanding of radiation reaction effects is of special interest. Here, we describe how the inclusion of quantum radiation reaction effects changes the dynamics of ultra-relativistic electron beams colliding with intense laser pulses significantly. Thereafter, the angular distribution of emitted radiation is demonstrated to be strongly altered in the quantum framework, if in addition to single photon emission also higher order photon emissions are considered. Furthermore, stimulated Raman scattering of an ultra-intense laser pulse in plasmas is examined and forward Raman scattering is found to be significantly increased by the 
inclusion of radiation reaction 
effects in the classical regime. The numerical simulations in this work show the feasibility of an experimental verification of the predicted effects with presently available lasers and electron accelerators.
\end{abstract}

\section{Introduction}

If a charged particle, an electron for definiteness, is exposed to an electromagnetic background field, it will be accelerated and subsequently emit radiation. Although this process is fundamental in electrodynamics, the usual classical treatment is insufficient, since it does not take into account radiation reaction (RR), i.e., the back reaction of the emitted radiation on the charged particle itself \cite{Di_Piazza_2012,Rohrlich:2002}. In this work high intensity plane wave laser fields of electric field amplitude $\mathcal{E}_0$, central angular frequency $\omega_0$, central wavelength $\lambda_0$ and propagation direction $\bm{n}_0$ are investigated, in order to theoretically probe the parameter regime relevant for RR. The study of electron-laser interactions in classical as well as quantum electrodynamics (QED) is of special interest, as they are expected to have impact on various fields like accelerators, quantum x-free electron lasers \cite{Bonifacio_1984,Bonifacio_1985,Moshammer_2009} or the 
production of multi-GeV photon beams \cite{
Apyan_2005}. However, RR effects are also of pure theoretical interest, as even in the framework of classical electrodynamics theoretical methods such as renormalization are necessary to describe the self-coupled dynamics \cite{Landau_b_2_1975}. 

Recently it has been reported, that the so-called Landau-Lifshitz (LL) equation is in theory the accurate equation of motion for an electron of mass $m$ and charge $e<0$ in the framework of classical electrodynamics \cite{Di_Piazza_2012,Rohrlich:2002,Landau_b_2_1975,Spohn:2000,Spohn:2004,Rohrlich_b_2007,Bulanov2011a}. (Units with $\hbar=c=1$ are used.) In case of an electromagnetic plane wave the LL equation allows for an analytical solution \cite{DiPiazza:2008} and it was demonstrated that if the parameter $R_c=\alpha\chi_0\xi_0$ is of order of unity, the dynamics of an electron with initial momentum $p_i^\mu$ colliding with a plane wave laser field is significantly altered by RR effects. Here, we introduced the classical and quantum nonlinearity parameters $\xi_0=|e|\mathcal{E}_0/m\omega_0$ and $\chi_0=((n_0p_i)/m)\mathcal{E}_0/E_{cr}$, respectively, and defined $n_0^{\mu}=k_0^\mu/\omega_0=(1,\bm{n}_0)$, where we made use of the abbreviation  $(ab) = a_\mu b^\mu$ denoting the product of two four-vectors 
$a^\mu$ and $b^\mu$. 
Furthermore, $\alpha=e^2$ is the fine structure constant and $E_{cr}=m^2/|e|=1.3\times10^{16}\, \textnormal{V}/\textnormal{cm}$ is the critical field of QED. In addition, in a bichromatic laser pulse the RR force was shown to modify the trajectory of an electron also in the case of $R_c\ll1$ \cite{Tamburini_2013} (see also \cite{Di_Piazza_2009}). The head-on collision of a laser field with an ultra-relativistic electron of initial energy $\varepsilon_i$ yields $R_c=3.2\,\varepsilon_i[\textnormal{GeV}]I_0[10^{23}\;\textnormal{W/cm$^2$} ] / \omega_0[\textnormal{eV}] $, where $I_0=\mathcal{E}_0^2/4\pi$ is the laser peak intensity. Thus, the experimental challenges in observing RR effects can be understood, since this expression is usually very small. Nevertheless, an alternative method has been proposed to measure RR effects also at moderate laser intensities and larger pulse durations \cite{Heinzl2013}. However, for upcoming ultra-high intensity laser facilities it was calculated that for laser intensities 
exceeding $I_0 > 2\times 10^{23}\, \textnormal{W/cm$^2$}$ RR effects have to be taken into account and for $I_0 > 4\times 10^{24}\, \textnormal{W/cm$^2$}$ quantum effects will become important \cite{Bulanov2011} and lead to a strong alteration of the particle's dynamics \cite{Zhidkov_2013}.

In addition, the consistency of QED and the diverse classical approaches was examined \cite{Ilderton2013}. In the quantum description a laser field is depicted as a stream of photons and the scattering process with an electron leads to the absorption of many photons from the field and to the subsequent emission of one or more photons. In fact, the emission of a single photon by an electron in strong laser pulses (nonlinear single Compton scattering (NSCS)) has been studied thoroughly \cite{Boca2009,Mackenroth2011,Seipt2011,Krajewska2012,Harvey_2012} and the classical spectra without RR \cite{Jackson_b_1975} were shown to coincide with the NSCS spectra for $\chi_0\ll1$. Hence,  in the ultra-relativistic regime and for negligible pair creation the quantum analogue of RR can be understood as the emission of a higher number of photons \cite{DiPiazza:2010mv}. Providing the first results for such higher order photon emission, the emission of two photons by an 
electron in a plane wave (nonlinear double Compton scattering (NDCS)) was considered recently \cite{Lotstedt2009a,Lotstedt2009b,Seipt2012,Mackenroth2013}.

The influence of the RR force on the collective particle dynamics of a system can be considerably different from the single particle dynamics as the collective energy loss of the particles due to RR can give rise to unexpected physical phenomena in a medium. In the classical electrodynamics regime, an analysis of the influence of the Landau-Lifshitz RR force \cite {Landau_b_2_1975} on the collective plasma dynamics of the particles has been carried out recently, where it was found that the inclusion of RR counterintuitively strongly enhances the forward Raman scattering (FRS) of the laser radiation in plasmas \cite{kumar13}. This growth enhancement is attributed to the nonlinear mixing of the two Raman sidebands mediated by the RR force.

In the sections concerning quantum RR (sections 2 and 3), we will employ light-cone coordinates, which for a four-vector $a^{\mu}=(a^0,\bm{a})$ are defined as $a^{\mu}=(a^+,a^{\textnormal{\textminus}},\bm{a}^{\perp})$, where $a^{\pm}=a^0\pm a^{\|}$, with $a^{\|}=\bm{k}_0\cdot\bm{a}/\omega_0$ and $\bm{a}^{\perp} = \bm{a}-a^{\|}\bm{k}_0/\omega_0$.

\section{Kinetic approach to quantum radiation reaction}

In this section, we investigate RR effects in the collision of an intense laser pulse with an ultra-relativistic electron beam. In classical electrodynamics RR was shown to reduce the energy width of electron \cite{Zhidkov_2002} and ion \cite{Naumova_2009,Chen_2010,Tamburini_2010,Tamburini_2011} bunches. However, in the quantum regime we find that RR has the opposite tendency and leads to a broadening in the width of the energy distribution describing the electron beam. The difference between the classical and the quantum regime can be explained by the increasing importance of the stochastic nature of photon emission in the quantum regime. The classical LL equation ignores the stochasticity of photon emission and even for small $\chi_0$ a correct treatment of RR requires an additional stochastic term. In fact, a Langevin-like equation can be employed for not too large $\chi_0$'s, though in the full quantum regime at $\chi_0\sim1$  this approximated description is not valid anymore. The broadening in the energy 
distribution of the electron beam displayed by our 
numerical simulations is expected to be detectable with nowadays available electron accelerators and strong laser fields.

An exact treatment of RR in the realm of strong-field QED would in principal result in the determination of the full $S$-matrix, taking into account multiple photon emission, radiative corrections and pair creation following photon emission \cite{Di_Piazza_2012,DiPiazza:2010mv}. However, RR mainly stems from incoherent multi-photon emission in the so-called ``nonlinear moderately-quantum'' regime $\xi_0\gg 1,\ \chi_0\lesssim 1$ \cite{DiPiazza:2010mv}, where nonlinear effects in the laser field are considered to be large and quantum effects are already important but pair production can still be neglected. In order to investigate RR in this regime we apply a kinetic approach \cite{Baier_b_1998,Khokonov_2004,Sokolov:2010am} and in turn characterize electrons and photons by distribution functions. Due to the neglect of pair production the distribution function of positrons is considered to vanish and the kinetic equation of the electrons is decoupled from that of the photons \cite{Baier_b_1998,Khokonov_2004,
Sokolov:2010am}. If the average energy of the electron beam $\varepsilon^*$ fulfills the constraint $\varepsilon^*\gg m\xi_0$, the transverse momentum of the electrons can be disregarded, since throughout the whole interaction it will remain much smaller than the longitudinal momentum \cite{Landau_b_2_1975}. Assuming the collision of presently available optical ($\omega_0=1.55\;\textnormal{eV}$) laser fields of intensity $10^{22}\;\textnormal{W/cm$^2$}$ \cite{Yanovsky_2008} with electron bunches with typical energies of $\varepsilon^*=1\;\textnormal{GeV}$ yields $m\xi_0=25\;\textnormal{MeV}$ and allows us to treat the present problem as an one-dimensional one.
Considering a linearly polarized plane wave propagating along the positive $y$ direction, we introduce the electric field of the laser field by $\bm{\mathcal{E}}(\eta)=\mathcal{E}_0g(\eta)\hat{\bm{z}}$ depending on the laser phase $\eta=\omega_0(t-y)$ via a pulse-shape function with $|g(\eta)|_{\textnormal{max}}\le 1$. Since the ultra-relativistic regime $\xi_0\gg1$ is investigated, the emission in a plane wave field of a photon with four-momentum $k^{\mu}=(\omega,\bm{k})$ by an electron with initial four-momentum $p^{\mu}=(\varepsilon,\bm{p})$ can be characterized by adopting  the well-known single photon differential emission probability $dP_{p^-}$ per unit phase and per unit $u=k^-/(p^--k^-)$ \cite{Ritus1985}
\begin{equation}
\hspace*{-2cm}\label{ToniProb}
\frac{dP_{p^-}}{d\eta du}=\frac{\alpha}{\sqrt{3}\pi}\frac{m^2}{\omega_0p^-}\frac{1}{(1+u)^2}\left[\left(1+u+\frac{1}{1+u}  \right) \textnormal{K}_{\frac{2}{3}}\left( \frac{2u}{3\chi}\right)-\int_\frac{2u}{3\chi}^\infty dx\, \textnormal{K}_{\frac{1}{3}}(x) \right],
\end{equation}
where $\textnormal{K}_\nu(\cdot)$ is the modified Bessel function of $\nu$th 
order. The quantum nonlinearity parameter is given by $\chi\equiv\chi(\eta,p^-)=(p^-/m)|\mathcal{E}(\eta)|/E_{\textnormal{cr}}$ and is now depending on the laser phase due to the oscillating electric field $\mathcal{E}(\eta)=\mathcal{E}_0g(\eta)$. The fact that Eq. (\ref{ToniProb}) only depends on the variables $\eta$ and $p^-$ allows us to employ an electron distribution $f_e(\eta,p^-)$ for the description of an electron beam and its phase evolution is determined by the kinetic equation (see Ref. \cite{Baier_b_1998})
\begin{equation}
\label{Kinetic}
\frac{\partial f_e}{\partial\eta}=\int_{p^-}^\infty dp'^-\,\frac{dP_{p'^-}}{d\eta dp^-}f_e(\eta,p'^-)-f_e(\eta,p^-)\int_0^{p^-}dk^-\frac{dP_{p^-}}{d\eta dk^-}
\end{equation}
with
\begin{eqnarray}
\frac{dP_{p'^-}}{d\eta dp^-}&=\frac{p'^-}{(p^-)^2} \left.\frac{dP_{p'^-}}{d\eta du}\right\vert_{u=\frac{p'^--p^-}{p^-}},\\
\frac{dP_{p^-}}{d\eta dk^-}&=\frac{p^-}{(p^--k^-)^2} \left.\frac{dP_{p^-}}{d\eta du}\right\vert_{u=\frac{k^-}{p^--k^-}}.
\end{eqnarray}
Eq. (\ref{Kinetic}) is non-local in the momentum $p^-$ since it is an integro-differential equation. Hence, an electron with initial momentum $p_0^-$ is coupled to the electron with momentum $p_0^--k^-$, where the momentum of the emitted photon $k^-$ varies between $0$ and $p_0^-$. In turn, the evolution of $f_e(\eta,p^-)$ is not only influenced by the neighborhood of $p^-$ but by all possible values of $p'^-$. In order to study the classical limit of RR, we expand Eq. (\ref{Kinetic}) up to the order of $\chi^3(\eta,p^-)$
\begin{equation}
\label{FP}
\frac{\partial f_e}{\partial\eta}=-\frac{\partial}{\partial p^-}[A(\eta,p^-)f_e]
+\frac{1}{2}\frac{\partial^2}{\partial (p^-)^2}\left[B(\eta,p^-)f_e\right],
\end{equation}
which is a Fokker-Planck-like equation \cite{Sokolov:2010am,Gardiner_b_2009,Lifshitz_1981}. Here, we introduced the ``drift'' coefficient $A(\eta,p^-)=-\frac{2\alpha m^2}{3\omega_0}\chi^2(\eta,p^-)[1-\frac{55\sqrt{3}}{16}\chi(\eta,p^-)]$ and the ``diffusion'' coefficient $B(\eta,p^-)=\frac{\alpha m^2}{3\omega_0}\frac{55}{8\sqrt{3}}p^-\chi^3(\eta,p^-)$. Eq. (\ref{FP}) is no longer an integro-differential equation and the evolution of $f_e(\eta,p^-)$ is determined by the momenta $p'^-$ close to $p^-$, due to the locality in $p^-$. If also higher-order corrections in $\chi(\eta,p^-)$ are considered, the expansion of Eq. (\ref{Kinetic}) leads to terms with higher derivatives of $f_e(\eta,p^-)$ with respect to $p^-$.

By expanding the kinetic equation (\ref{Kinetic}) we obtain two quantum corrections to the classical kinetic equation $\partial f_e/\partial\eta=-\partial/\partial p^-\left(f_e d p^-/d\eta\right)$, with $d p^-/d\eta=-I_{cl}(\eta,p^-)/\omega_0$ and the classical radiation intensity $I_{cl}(\eta,p^-)=(2/3)\alpha m^2 \chi^2(\eta,p^-)$. The first correction modifies the drift coefficient $A(\eta,p^-)$, but does not affect the analytical structure of the classical equation. Therefore the phase evolution is purely deterministic \cite{Neitz2013} and this correction coincides with the well-known leading-order correction 
to the total intensity of radiation \cite{Baier_b_1998,Ritus1985}. As the correction term is negative, we expect the reduction of the energy width to be smaller than in classical electrodynamics. However, if the classical radiation intensity $I_{cl}(\eta,p^-)$ is substituted by the corresponding quantum one $I_q(\eta,p^-)$ (see, e.g., \cite{Ritus1985}), the corresponding Liouville equation still predicts a decrease in the energy width, due to the fact that electrons with higher energy on average will emit more radiation. On the other hand, the second quantum correction introduces the diffusion coefficient $B(\eta,p^-)$ and transforms  the classical kinetic equation into a Fokker-Planck-like equation, which is equivalent to the single-particle stochastic equation $dp^-=-A(\eta,p^-)d\eta+\sqrt{B(\eta,p^-)}dW$, with $dW$ an infinitesimal stochastic function \cite{Gardiner_b_2009}. Reflecting the stochasticity of photon emission, this equation is no longer deterministic and the stochastic evolution of the 
system causes the broadening of the energy distribution \cite{Neitz2013}.

\begin{figure}
\includegraphics[width=\linewidth]{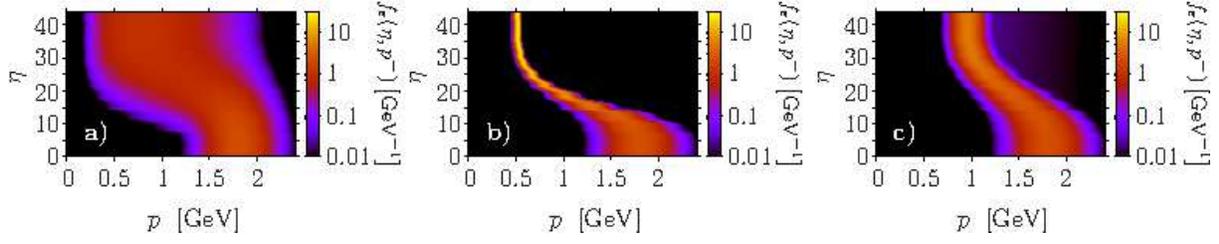}
\caption{(color online) Phase evolution of the electron distribution for a 7-cycle $\sin^2$-like laser pulse according to Eq. (\ref{Kinetic}) (part a)), to the classical kinetic equation without (part b)) and with the replacement $I_{cl}(\eta,p^-)\rightarrow I_q(\eta,p^-)$ (part c)). The laser and the initial electron distribution parameters are given in the text.}
\label{RRquant}
\end{figure}

We solved Eq. (\ref{Kinetic}) numerically by employing a finite difference method and we ensured that for $\chi^\ast=(p^{*,-}/m)\mathcal{E}_0/E_{\textnormal{cr}}\ll1$, where $p^{*,-}$ is the average momentum, our numerical simulations coincide with classical results, as well as with the results in the quantum regime \cite{DiPiazza:2010mv}. We now consider a laser pulse with shape function $g(\eta)=\sin^2(\eta/2N_L)\sin(\eta)$, where $N_L$ is the number of laser cycles and with $\omega_0=1.55\;\textnormal{eV}$, to collide head-on with an initial Gaussian electron distribution that is normalized to unity. Further, we assume the laser peak intensity $I_0=2.5\times 10^{22}\;\textnormal{W/cm$^2$}$, the average momentum $p^{*,-}=1.8\;\textnormal{GeV}$ ($\varepsilon^*\approx 900\;\textnormal{MeV}$) and the initial width of the electron distribution $\sigma_{p^-}=0.18\;\textnormal{GeV}$ corresponding to $\chi^*=0.8$, and $N_L=7$ corresponding to about $21\;\textnormal{fs}$. The results of our numerical simulations 
are shown in Fig. \ref{RRquant}.
As expected, Fig. \ref{RRquant}a) displays a broadening of the electron distribution in the quantum regime. However, if the full kinetic Eq. (\ref{Kinetic}) is not applied but the quantum intensity $I_q(\eta,p^-)$ (see, e.g., \cite{Baier_b_1998,Ritus1985}) is set into the classical kinetic equation, the photon emission still reduces the energy spread of the electron distribution (Fig. \ref{RRquant}c)) as in  the classical case (Fig. \ref{RRquant}b)). This clearly indicates that the broadening of the electron distribution is induced by the importance of the stochasticity of photon emission in the full quantum regime.

\section{Nonlinear double Compton scattering}

In this section, we are going to demonstrate how in the ultra-relativistic quantum regime, distinguished by the conditions $\xi_0\gg1$, $\chi_0\gtrsim1$, the NDCS signal can be spatially separated from the dominant NSCS signal. The investigation of the given parameter regime is timely as it is about to come into experimental reach. To clearly interpret the results of this analysis and to establish an intuitive understanding of the underlying physics we are going to work out a semi-classical picture of two smoothly joined classical electron trajectories from which the two separate photons are consecutively emitted. These trajectories are found to feature a discontinuity only in the electron's energy, which can be attributed to the emission of a photon of finite energy. We are going to connect this picture of a discontinuous change of the trajectory to the classical account of RR which leads to a smooth change of the electron's trajectory \cite{DiPiazza:2010mv}. Finally, we are going to 
demonstrate that the discussed effect is likely to be observable with already available lasers, featuring intensities beyond $10^{22}\textnormal{ W}/\textnormal{cm}^2$ \cite{Yanovsky_2008}, and electron acceleration technology either from conventional accelerators \cite{ERL} or modern plasma-based laser accelerators \cite{Leemans_2006,Clayton_2010}.

As in the regime $m\xi_0 \ll \varepsilon_i$, with the electron's initial energy $\varepsilon_i$, the radius of the laser's focal volume routinely exceeds the electron's perpendicular excursion, which is on the order of $\lambda_0(m\xi_0/\varepsilon_i)$ it is justified to approximate the laser field by a plane wave. In the present study we are thus going to model the laser pulse by a plane wave field of arbitrary temporal shape $A_0^\mu (\eta) = (\mathcal{E}_0/\omega_0) \epsilon_0^{\mu}\psi(\eta)$ depending on the space-time coordinates only via the invariant phase $\eta=k_0^\mu x_\mu$. Here $\epsilon^{\mu}_0$ is the wave's polarization four-vector and the shape function $\psi(\eta)$ describes the laser pulse's arbitrary temporal shape. Since in the regime $\xi_0\gg1$ nonlinear effects have to be taken into account exactly we will perform our calculations in the so-called Furry picture of quantum dynamics \cite{Furry_1951,Fradkin_Gitman}. The essence of this procedure is to attribute terms in the QED 
Lagrangian describing the strong background field, i.e.\ the laser pulse in this case, to the free Lagrangian and to subsequently quantize the charged fermionic fields in the presence of this strong background. Technically this amounts to a replacement of the vacuum wave function of an electron with momentum $p^{\mu}$ and spin quantum number $\sigma$ by a solution of the Dirac equation in the presence of the strong plane wave background, known as Volkov wave function $\Psi_{p,\sigma}(x)$ \cite{Landau_b_4_1982}. It is then straightforward to obtain an expression for the scattering matrix element $S_{fi}$ of an electron with initial (final) four-momentum $p_i^{\mu}=(\varepsilon_i,\bm{p}_i)$ ($p_f^{\mu}=(\varepsilon_f,\bm{p}_f)$) and spin quantum number $\sigma_i$ ($\sigma_f$) emitting two photons with wave vectors $k^{\mu}_1$ and $k_2^{\mu}$ and polarization four-vectors $\epsilon^{\mu}_{k_1,\lambda_1}$ and $\epsilon^{\mu}_{k_2,\lambda_2}$, respectively. The resulting expression can be written as $S^{(1)}_{fi}
+S^{(2)}_{fi}$ with
\begin{eqnarray}
\hspace*{-1cm} S^{(1)}_{fi} =& - e^2\int \d^4 x \d^4 y\ \overline{\Psi}_{p_f,\sigma_f}(y)\,\epsSlash_{k_2,\lambda_2}^*\e^{\i k_2y}\,G(y,x)\,\epsSlash_{k_1,\lambda_1}^*\e^{\i k_1x}\,\Psi_{p_i,\sigma_i}(x) \label{Eq:Matrix_Element_int}
\end{eqnarray}
and $S^{(2)}_{fi} = S^{(1)}_{fi}(1\leftrightarrow 2)$. Here $/\!\!\!a=\gamma^{\mu}a_\mu$ is the common Feynman slash notation with the four-vector of the Dirac matrices $\gamma^\mu$. Furthermore, we introduced the Dirac conjugate wave function $\overline{\Psi}_{p,\sigma}(x)=\Psi^{\dag}_{p,\sigma}(x)\gamma^0$ and made use of the laser dressed electron propagator $G(y,x)$ \cite{Ritus1985}. Due to symmetry reasons it is sufficient to only compute the quantity $S^{(1)}_{fi}$, whence the cross-channel amplitude $S^{(2)}_{fi}$ can be obtained by the exchange of indices $(1\leftrightarrow2)$ in the final expression. According to earlier work \cite{Seipt2012,Mackenroth2013} this partial amplitude naturally splits up into two contributions in the following way
\begin{equation}
\hspace*{-2.1cm}   S^{(1)}_{fi} = (2\pi)^3\sum_{r,s=0}^2(a_rf_r\delta_{r,s}+b_{r,s}f_{r,s}) \delta(p_i^--k_1^--k_2^--p_f^-)\delta^{(2)}(\bm{p}_i^{\perp}-\bm{k}_1^{\perp}-\bm{k}_2^{\perp}-\bm{p}_f^{\perp}). \hskip .1cm \label{Eq:Amplitude_SplitUp}
\end{equation}
The matrix coefficients $a_r$ and $b_{r,s}$ are rather involved but not needed here and thus not given explicitly. Anyway, the important information on the dynamics of the two-photon emission process is encoded in the dynamic integrals \cite{Seipt2012,Mackenroth2013}
\begin{subequations}\label{Eq:DynamicIntegrals}
 \begin{align}
  f_r =& \int \d \eta \psi^r(\eta)\ \text{exp}\{-\i\,[S_x(\eta)+S_y(\eta)]\}, \label{Eq:DynamicIntegrals_fi}\\
  f_{r,s} =& \int \d \eta_x\d \eta_y \Theta(\eta_y-\eta_x)\ \psi^s(\eta_x)\psi^r(\eta_y) \text{exp}\{-\i\,[S_x(\eta_x)+S_y(\eta_y)]\}, \label{Eq:DynamicIntegrals_fij}
 \end{align}
\end{subequations}
where $S_{x/y}(\eta)=\int_0^{\eta}\d \eta'[\alpha_{x/y}\,\psi(\eta')+\beta_{x/y}\,\psi^2(\eta')+\gamma_{x/y}]$, with $\alpha_x = -m\xi[(p_i\epsilon_0)/(k_0p_i)-(p_t\epsilon_0)/(k_0p_t)]$, $\beta_x = -m^2\xi^2(k_0k_1)/2(k_0p_t)(k_0p_i)$, $\gamma_x = -(k_1p_i)/(k_0p_t)$. The parameters $\alpha_y$, $\beta_y$ and $\gamma_y$ are obtained from $\alpha_x$, $\beta_x$ and $\gamma_x$, respectively, substituting $p^{\mu}_t\to p^{\mu}_f$, $p^{\mu}_i\to p^{\mu}_t$ and $k^{\mu}_1\to k^{\mu}_2$. In these expressions we defined the transitional electron momentum $p_t^\mu$, distinguished by the four conditions $p_t^-=p_i^--k_1^-$, $\bm{p}_t^{\perp}= \bm{p}_i^{\perp} - \bm{k}_1^{\perp}$ and $p_t^2=m^2$. From the above expressions one can read off that the integrals $f_0$, $f_{0,s}$ and $f_{r,0}$ are divergent. These divergences, however, can be analytically regularized by an integration by parts technique \cite{Seipt2012,Mackenroth2013}. Carrying 
out the resulting replacements in the scattering matrix element, taking the modulus square and summing over the discrete and continuous phase spaces of all particles involved in the scattering the final result for the differential energy spectrum is found to be
\begin{equation}
\d E = \frac{\omega_1+\omega_2}{2}\frac{d^3\bm{p}_f}{(2\pi)^3}\prod_{i=1}^2\frac{d^3\bm{k}_i}{(2\pi)^3}\sum_{\{\sigma,\lambda\}}\Big|S^{(1)}_{fi}+S^{(2)}_{fi}\Big|^2 , \label{Eq:EnergySpectrum}
\end{equation}
where $\{\sigma,\lambda\}\equiv \sigma_i,\sigma_f,\lambda_1,\lambda_2$. The $\delta$-functions contained in $S^{(1,2)}_{fi}$ serve to fix the final electron momentum $p_f^\mu$ whence we only have to integrate over the phase spaces of the emitted photons.

As a next step we wish to give an exemplary numerical study of Eq.\ (\ref{Eq:EnergySpectrum}) to demonstrate its tractability as well as to highlight the aforementioned spatial separation of the NSCS and the NDCS emission signals. For the numerical evaluation we assume the laser's shape function to be $\psi(\eta)=\sin^4(\eta/4)\sin(\eta)$ for $\eta\in\left[0,4\pi\right]$ 
and zero elsewhere with a central frequency $\omega_0=1.55$ eV, corresponding to an optical laser pulse of approximately $5$ fs duration. We consider the case of an ultra-intense laser pulse of peak intensity $I_0=10^{22} \textnormal{W}/\textnormal{cm}^2$ corresponding to $\xi_0\approx48$, as is already available at nowadays working laser facilities such as the Hercules laser \cite{Yanovsky_2008}. Furthermore, we study the case of an ultra-relativistic electron of initial energy $\varepsilon_i=5$ GeV colliding head on with the specified laser pulse, resulting in a quantum nonlinearity parameter $\chi_0=2.8$, indicating the importance of quantum effects. Next, we wish to recall several properties of the NSCS signal, that are to be expected for the given experimental setup \cite{Mackenroth2010}: At $\xi_0 \gg 1$ the single photon signal is limited to a polar angular range $\pi-\theta\le \psi_0\,\theta_0$, where the typical opening angle of the single photon emission cone is given by $\theta_0=m\xi_0/\varepsilon_i$. The maximum value of the shape function specified above is $\psi_0=|\max(\psi(\eta))|=0.8$. Any emission predicted towards polar angles outside this emission cone will be clearly separated from the dominant NSCS signal.

As our reference frame we choose the coordinate frame in which the laser pulse is polarized along the $x$-direction and propagates towards the positive $z$-axis. This axis we also choose as polar axis, i.e.\ $\theta=\pi$ corresponds to the electron's initial propagation direction opposite to the laser's propagation direction at $\theta=0$. To stress that two-photon emission will be important for the electron's radiation pattern given the specified experimental parameters we estimate the average number of emitted photons to be $\mathcal{N}_{NSCS}=1.6$ \cite{Ritus1985}. 

\begin{figure}
\centering
\includegraphics[width=\linewidth]{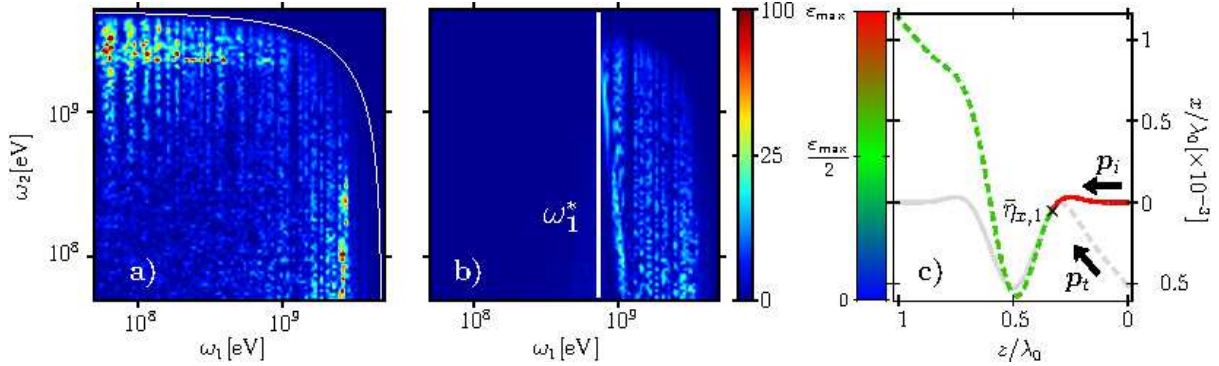}
\caption{(color online) Two-photon energy emission spectra $\d E/\Pi_{i=1}^2\d \omega_i\d\Omega_i [\textnormal{eV$^{-1}$ sr$^{-2}$}]$ at $\chi_0\approx 2.8$ observed at $\theta_1=\pi-\theta_0/2$, and at $\theta_2=\theta_1$ (part a)) and at $\theta_2=\pi-\theta_0$ (part b)), with $\theta_0=5\times 10^{-3}\;\textnormal{rad}$. Other numerical parameters are given in the text. The solid white lines correspond to the cutoff-energy $\omega_1+\omega_2=\varepsilon_i$ (part a)) and the threshold frequency $\omega_1^*$ (part b)), respectively. Part c): Two classical electron trajectories with initial electron momentum $\bm{p}_i$ (solid line) and $\bm{p}_t$ (dashed line). In color-code: Actual electron trajectory for a photon with energy $\omega_1=2$ GeV emitted at $\bar{\eta}_{x,1}$ towards $(\theta_1,\phi_1)$.}
 \label{fig:Recoil_Spectra}
\end{figure}

In the given reference frame we consider an experiment where one photon detector observes any photon emitted towards the direction ($\theta_1=\pi-\theta_0/2,\phi_1=\pi$) and a second detector to trace photons emitted towards the two different directions ($\theta_2=\pi-\theta_0/2,\phi_2=0$) (see fig.\ \ref{fig:Recoil_Spectra}a)) and ($\theta_2=\pi-\theta_0,\phi_2=0$) (see fig.\ \ref{fig:Recoil_Spectra}b)). While the first detector is positioned inside the NSCS emission cone the choice for the second observation direction corresponds to photon detection inside  (see fig.\ \ref{fig:Recoil_Spectra}a)) and outside (see fig.\ \ref{fig:Recoil_Spectra}b)) this cone, respectively. The azimuthal angles $\phi_{1,2}=0,\pi$ are chosen for observation of the emitted photons within the laser's plane of polarization, where most radiation is emitted.

In fig.\ \ref{fig:Recoil_Spectra}a) we wish to highlight that the cutoff for the emitted photon's frequencies (white line in fig.\ \ref{fig:Recoil_Spectra}a)), set by energy-momentum-conservation, is closely approached, whence we infer that quantum effects indeed are non-negligible. Comparing now this figure to Fig.\ \ref{fig:Recoil_Spectra}b), which is plotted in the same color-scale, we immediately conclude that there is a considerable amount of radiation emitted to directions outside the NSCS emission cone (recall that this latter figure corresponds to the detection of a photon outside this cone). We can thus conclude a clear spatial separation between the NSCS signal, confined exclusively to the angular range $\pi-\theta\le 0.8\,\theta_0$, and the NDCS signal, which is also detectable under $\pi-\theta\ = \theta_0$. The classical analog of this quantum result is an effect attributed to RR changing the angular distribution of the radiation emitted by an electron \cite{Di_Piazza_2009}. For a qualitative 
interpretation of the presented separation of the single and two photon signals we take advantage of a stationary-phase analysis that was recently developed for the analysis of NDCS spectra \cite{Mackenroth2013}. It was shown that it was only necessary to analyze the part of the scattering matrix element proportional to the bivariate dynamic integrals $f_{r,s}$ as this partial amplitude largely dominates the scattering amplitude. The corresponding dynamic integrals were approximated as \cite{Mackenroth2013} $f_{r,s} \approx \sum_{l,n} \Theta(\bar{\eta}_{y,n}- \bar{\eta}_{x,l}) f^y_r(\bar{\eta}_{y,n}) f^x_s(\bar{\eta}_{x,l})$ with $f^{x/y}_r = \int \d \eta \psi^r(\eta)\ \textnormal{exp}[-\i S_{x/y}(\eta)]$ with the exponential phases $S_{x/y}(\eta)$ defined above. The sum over the indices $l$ and $n$ is a sum over all stationary points in whose vicinities the only non-negligible contributions to the dynamic integrals are formed. These stationary points are found as solutions of the equations $\psi\left(\bar{\eta}_{x,l}\right) = -1/2$ and $\psi(\bar{\eta}_{y,n}) = \Delta\vartheta_2 - \omega_1/\varepsilon_i\left( 1/2 + \Delta\vartheta_2 \right)$ with $\Delta\vartheta_2=(\pi-\theta_2)/\theta_0$. In these expressions the fixed values of $\theta_1, \phi_1$ and $\phi_2$ were already inserted and only $\theta_2$ was left variable. The given equations formally correspond to the stationary point conditions of two separate single photon emissions, where first a photon with wave vector $k_1^\mu$ is emitted by an electron with initial momentum $p_i^\mu$, whereas the second photon with wave vector $k_2^\mu$ is emitted by an electron with initial momentum $p_t^\mu$. We can consequently interpret two-photon emission in the regime $\xi_0\gg1$ as the sequential emission of the first and second photon emitted from the classical trajectory an electron would take if it entered the laser field with an initial four-momentum $p^{\mu}_i$ and $p^{\mu}_t$, respectively. For the sake of continuity these trajectories naturally have to be 
joined at the emission point of the first photon $k^{\mu}_1$. However, the energies a classical electron will have when following these two separate trajectories will be discontinuous at the connection point, reflecting the finite loss of energy and momentum due to the quantum photon emission. We note that, of course, there are several solutions for the stationary points $\bar{\eta}_{x,l}$ and hence several combinations of classical trajectories for each choice of observation directions for the emitted photons. One exemplary combination of two such classical trajectories computed for an electron scattered from a laser pulse of the given parameters is shown in fig.\ \ref{fig:Recoil_Spectra}c). An electron with initial momentum $p_i^\mu$ would follow the solid trajectory, whereas, on the other hand, an electron of initial electron momentum $p_t^\mu$ would take the dashed trajectory. In the case shown here the transitional momentum $p_t^\mu$ is computed from $p_i^\mu$ by assuming the emission of a photon of 
energy $\omega_1=2$ GeV into the direction ($\theta_1,\phi_1$). Since this corresponds to the loss of a significant portion of its initial energy by the electron the momentum $p_t^\mu$ significantly differs from $p_i^\mu$ and thus the two corresponding trajectories are clearly distinguishable. This feature then also explains the considerable spatial separation of the NSCS and the NDCS signal in the regime $\chi_0\sim1$. We note, however, that the tangent vectors of the two shown trajectories are parallel at the junction, as they have to be for an ultra-relativistic electron emits photons almost exclusively into its instantaneous direction of propagation and thus can lose momentum only in this direction. The aforementioned loss of energy, however, can be read off from the color-coding of the two trajectories and is clearly discontinuous at the point of emission of the first photon. It is also this significantly decreased energy of the electron that leads to a stronger 
deflection of the second trajectory inside the laser field. To find an analytical prediction of the energy threshold the first photon has to carry away to render this deflection strong enough to facilitate emission towards the chosen observation direction $\theta_2 = \pi-\theta_0$ we solve the defining equation of the stationary point $\bar{\eta}_{y,n}$ for the first emitted photon's frequency $\omega_1$. From this procedure we find the threshold $\omega_1^*= \varepsilon_i (1 - 0.8)/(1 + 1/2)\approx 660$ MeV, which is well confirmed in fig.\ \ref{fig:Recoil_Spectra}b). Furthermore, we have to stress that in the overall scattering amplitude we also have to include the cross-channel term $S_{fi}^{(2)}$ where the photons' wave vectors $k_1^\mu$ and $k_2^\mu$ are exchanged. The interpretation for this cross-channel is in terms of classical trajectories is analogous to the above given arguments, however, with the order of the emission of the two photons exchanged as well.

\section{Influence of RR on the parametric instabilities in plasmas}

Parametric instabilities of a laser pulse in a plasma are important due to their applications in the area of laser-driven fusion, laser wakefield acceleration, and have been investigated for decades \cite{Kruer:2003yq,Brueckner:1974uq,drake:778,tripathi:468,mckinstrie:2626,Sakharov:1994fk,decker:2047,Barr:1999ve,jr.:1440,guerin:2807,Quesnel:1997fk,Esarey:2009fk}. The FRS---a scattering process belonging to the general Stimulated Raman scattering (SRS) processes in plasmas---is one of the prominent examples of parametric instabilities in plasmas. In the FRS, the incident pump laser decays into two forward moving daughter electromagnetic waves, and a plasma wave. The daughter waves have their frequencies upshifted (anti-Stokes waves) and downshifted (Stokes wave) from the pump laser by the magnitude which equals the excited plasma wave frequency.

Though, at high laser intensities $I_{0}\ge 10^{19}\,\text{W/cm}^2$, the growth rate of the parametric instabilities becomes smaller due to the relativistic Lorentz factor \cite{decker:2047}, the role of RR force becomes also important especially at ultra-high laser intensities, $I_{0}\ge 10^{22}\,\text{W/cm}^2$\cite{Di_Piazza_2012,Chen_2010,Tamburini_2010,Di_Piazza_2009,Schlegel:2012bh,Sokolov:2010kl,Keitel:1998fk}. Such ultra-intense laser systems are expected to be available in a near future after the commissioning of the Extreme Light Infrastructure (ELI) project in Europe \cite{eli:kx}. Due to the RR force, the laser pulse suffers damping while propagating in a plasma. As the laser loses energy due to RR force it facilitates, apart from the usual parametric decay processes, the availability of an additional source of free energy for perturbations to grow in the plasma. Its 
effective intensity also decreases which lowers the relativistic Lorentz factor. Moreover, the phase shift, caused by the RR force, in the nonlinear current densities causes polarization rotations of the scattered daughter electromagnetic waves. This necessitates to include the effect of RR force in the theoretical formalism of the parametric instabilities in the plasma.

We study the FRS of an ultra-intense laser pulse in a plasma including the RR force effects in the classical electrodynamics regime where quantum effects arising due to photon recoil and spin are negligible \cite{Di_Piazza_2012}. This approach is valid if the wavelength and magnitude of the external electromagnetic field in the instantaneous rest frame of the electron satisfy  $\lambda \gg \lambda_C,\, \mathcal{E} \ll {E}_{\text{cr}}$, where $\lambda_C = 3.9 \times 10^{-11}$ cm is the Compton wavelength and $E_{cr}$ is the critical field of the quantum electrodynamics \cite{Di_Piazza_2012}. For the laser intensities planned in the ELI project $I_{0}\sim 10^{22-23}\,\text{W/cm}^2$ \cite{eli:kx}, these two criteria can be fulfilled. In the classical electrodynamics regime, the Landau-Lifshitz RR force \cite{Landau_b_2_1975} correctly accounts for the radiation emitted by a relativistic 
charged particle \cite {Di_Piazza_2012}. 
Consider the propagation of a circularly polarized (CP) pump laser along the $\hat{\bm{z}}$ direction in an underdense plasma with uniform plasma electron density $n_e$. Ions are assumed to be at rest. Eq. of motion for an electron including the leading order term of the Landau-Lifshitz RR force in the laser field is 
\begin{equation}
\frac{\partial \bm{p}}{\partial t} + \bm{\upsilon}\cdot \nabla\bm{p}=-e \big(\bm{E}+\bm{\upsilon}\times\bm{B}\big) -\frac{2e^4}{3m^2}\gamma^2\bm{\upsilon}\Big[\big(\bm{E}+\bm{\upsilon}\times\bm{B}\big)^2  -\big({\bm{\upsilon}}\cdot \bm{E}\big)^2\Big],
\label{eom}
\end{equation} 
\noindent
where $\gamma=1/\sqrt{1-\upsilon^2}$ and the velocity of the light in vacuum $c=1$. The other terms of the Landau-Lifshitz RR force are ignored as they are $1/\gamma$ times smaller than the leading order term \cite {Landau_b_2_1975}.
We first ignore the RR term and express the electric and magnetic fields in potentials as $\bm{E}=-\nabla\phi-\partial\bm{A}/ \partial  t,\, \bm{B}=\nabla\times\bm{A}$. In a 1D approximation valid when $r_0\gg\lambda_0$ (where $r_0$ is the spot-size and $\lambda_0$ is the wavelength of the pump laser pulse), the transverse momentum and $z$-component of motion are $\bm{p}_{\perp}={e}\bm{A},\quad \text{and}\quad {\partial \upsilon_z}/{\partial t}={e\nabla \phi}/(m\gamma_0) -e^2\nabla |{A}|^2/(2m^2\gamma_0^2)$, where $\, \bm{A}=\bm{A}_0 e^{i\eta_{0}}/2 + c.c,\,\bm{A}_0=\bm{\sigma}A_{0},\bm{\sigma}=(\hat{\bm x}+i\hat{\bm y})/\sqrt{2},\gamma_0=(1+\xi_0^2/2)^{1/2},\xi_0=eA_0/m,\eta_0=k_0 z-\omega_0 t$, and $\omega_0,\,k_0$ are the carrier frequency and wavevector of the pump laser respectively \cite{decker:2047,Gibbon:2005ys}. A plane monochromatic CP light doesn't cause any charge separation since for it $\nabla |A_0|^2=0$ and consequently there is no 
component of velocity in the $\hat{z}$ direction. This solution is known as the Akhiezer-Polovin solution for a purely transverse monochromatic CP light in plasmas  \cite{decker:2047,Akhiezer:1956mz,S.V.-Bulanov:2001kx}. The scattering of the laser pulse results into the total vector potential of the form $\bm{A}=[\bm{A}_0 e^{i\eta_{0}}+\bm{\delta A}_{+}e^{i \bm{k}_{\perp}.\bm{x}_{\perp}} e^{i\eta_{+}} + \bm{\delta A}_{-}^{*}e^{-i\bm{k}_{\perp}.\bm{x}_{\perp}}e^{-i\eta^{*}_{-}}]/2 +c.c.$, where $\bm{\delta A}_{+}= \bm{\sigma} \delta A_{+}$, $\bm{\delta A}_{-}^{*}=\bm{\sigma} \delta A_{-}^{*}$, $\bm{\delta A}_{+}$ and $\bm{\delta A}_{-}$ represent the anti-Stokes and the Stokes waves respectively, $\eta_{+}=(k_z+k_0)z-(\omega+\omega_0)t,\,\eta_{-}^{*}=(k_z-k_0)z-(\omega^{*}-\omega_0)t$ \cite{decker:2047,Gibbon:2005ys}. Beating of the Stokes and the anti-Stokes waves with the pump laser leads to the plasma wave excitation $\delta n/ n_e$, which can be estimated from the equation of continuity, Poisson 
equation, and the $z$-component of equation of motion, yielding $\delta \tilde{n}=\left(e^2 k_z^2/2m^2\gamma_0^2 D_e\right)\left({A}_{0}^{*}{\delta A}_{+}+{A}_{0}{\delta A}_{-}\right)$, where $D_e=\omega^2-\omega_p^{'2},\, \omega_p^{'2}=\omega_p^{2}/\gamma_0,\, \omega_p^{2}=4\pi n_e e^2/m,\, \delta n/n_e=\delta \tilde{n}e^{i\eta}e^{i\bm{k}_{\perp}.\bm{x}_{\perp}}/2+c.c,$ and $\eta\equiv \eta_{+}-\eta_0 \equiv \eta_{-}+\eta_0=k_z z -\omega t$ \cite{decker:2047,Gibbon:2005ys}. Plasma wave oscillation causes an axial component of velocity and momentum $\upsilon_z \ll 1, p_z \ll p_{\perp}$. 

On using the above solutions for transverse and longitudinal components of momenta to simplify the RR term in Eq.\eqref{eom}, the full equation of motion after expressing the CP laser pulse as $\bm{A}=\bm{A_{\perp}}(\bm{x}_{\perp},z,t)e^{i\eta_0}/2+c.c.$, with its amplitude varying slowly i.e. $ \left|\partial \bm{A}_{\perp}/{\partial t}\right|\ll\left|\omega_0 \bm{A}_{\perp}\right|,\, \left|\partial \bm{A}_{\perp}/{\partial z}\right|\ll\left|k_0 \bm{A}_{\perp}\right|$, and $|\phi| \ll |\bm{A}|, \omega_p^2/\gamma \omega_0^2 \ll 1$, and $\gamma=(1+e^2|\bm{A}|^2/m^2)^{1/2}$, yields,   
\begin{gather}
\frac{\partial}{\partial t}\left(\bm{p}_{\perp}-{e}\bm{A}\right)=-{e\mu\omega_0}\bm{A}\gamma |\bm{A}|^2(1-2\upsilon_z),
\label{qmom}
\end{gather}
where $\mu=2e^4\omega_0/3m^3$, $\upsilon_z=(\omega/k_z)\,\delta \tilde{n}\, e^{i \bm{k}_{\perp}.\bm{x}_{\perp}} e^{i\eta}/2+c.c.$, and we have assumed $|\mu \gamma |\bm{A}|^2| \ll 1$, valid for laser intensity $I_{0}\le 10^{23}\,\text{W/cm}^2$. Since $|\phi| \ll |A|$ and the RR effects in the case of the collinear movement of plasma electrons and the plasma wave are negligible, we don't consider the effect of RR on plasma oscillations. One can solve Eq.\eqref{qmom} by expressing the transverse component of the quiver momentum as \emph{e.g.} $\bm{p}_{\perp}=[\bm{p}_0 e^{i\eta_{0}}+\bm{p}_{+}e^{i \bm{k}_{\perp}.\bm{x}_{\perp}} e^{i\eta_{+}} + \bm{p}_{-}^{*}e^{-i\bm{k}_{\perp}.\bm{x}
_{\perp}}e^{-i\eta^{*}_{-}}]/2+c.c.$, where $\bm{p}_{+}$ and $\bm{p}_{-}$ have similar polarizations as the anti-Stokes and the Stokes modes. The wave equation for the vector potential after the density perturbation $n=n_e+\delta n$ reads as
\begin{equation}
\nabla^2\bm{A}- \frac{\partial^2 \bm{A}}{\partial t^2}=\frac{\omega_p^{2}}{\gamma}\bigg( 1+\frac{\delta n}{n_e} \bigg)\frac{\bm{p}_{\perp}}{e}.
\label{waveeq}
\end{equation}

The dispersion relation for the equilibrium vector potential can be obtained after collecting the terms containing $e^{i\eta_0}$, and it reads as $\omega_0^2=k_0^2+{\omega_p^{'2}}\left(1-{i\mu}|A_0|^2\gamma_0/2\right)$, implying that the RR term causes damping of the pump laser field. We incorporate this damping by defining a frequency or a wavenumber shift in the pump laser \footnote{One can also incorporate the RR term by appropriately modifying the plasma frequency, which essentially implies change in the laser pump wavevector arising due to the it's dispersion in the plasma.}. On writing $\omega_0=\omega_{0r}-i\delta\omega_0,\,\delta\omega_0 \ll \omega_{0r}$ we get the frequency shift $\delta\omega_0$ as $\delta\omega_0={\omega_{p}^{'2}\varepsilon_{\text{rad}}\gamma_0 \xi_0^2}/{2\omega_{0r}}$, where $\varepsilon_{\text{rad}}=r_e\omega_{0r}/3,\,r_e=e^2/m$ is the classical radius of the electron and 
without the loss of generality we have assumed $\xi_0=\xi_0^*$. For the SRS growth to occur, this frequency shift must be less than the growth rate. On  collecting the terms containing $e^{ i\eta_{\pm}}e^{i\bm{k}_{\perp}.\bm{x}_{\perp}}$ in Eq.\eqref{waveeq}, we get  $D_{+}{{\delta A}_{+}}=R_{+}\left({{\delta A}_{+}}+{{\delta A}_{-}}\right)$ and $D_{-}{{\delta A}_{-}}=R_{-}\left({{\delta A}_{+}}+{{\delta A}_{-}}\right)$ which yields the dispersion relation
\begin{equation}
 \left(\frac{R_{+}}{D_{+}}+\frac{R_{-}}{D_{-}}\right)=1,
 \label{disp_rel}
\end{equation} 
where
\begin{gather}
D_{\pm}=(\omega \pm \omega_0)^2-\omega_{p}^{'2}\bigg(1-\frac{i\varepsilon_{\text{rad}} \xi_0^2\gamma_0\omega_0}{\omega\pm\omega_0}\bigg)-[(k_z\pm k_0)^2+k_{\perp}^2],\nonumber \\
R_{\pm}=\frac{\omega_p^2 \xi_0^2}{4\gamma_0^3}\Bigg[\frac{k_z^2 }{D_e}\left(1\mp i\varepsilon_{\text{rad}} \xi_0^2\gamma_0 + \frac{2 i \varepsilon_{\text{rad}} \xi_0^2\gamma_0}{k_z } \frac{\omega\omega_0}{\omega\pm\omega_0}\right) \nonumber  -\left(1 \mp i \varepsilon_{\text{rad}} \xi_0^2\gamma_0 \frac{\omega}{\omega\pm \omega_0}+ 4 i \varepsilon_{\text{rad}} \gamma_0^3\frac{\omega_0}{\omega\pm\omega_0}\right)\Bigg].
\label{pertAmpl}
\end{gather}
The RR term modifies the coupling between the Stokes and the anti-Stokes modes $(R_+ \neq R_-)$, and the form of dispersion relation from the dispersion relation derived before \cite{Kruer:2003yq,drake:778,decker:2047,Gibbon:2005ys}.

\begin{figure}
\centering
\includegraphics[width=0.5\textwidth,height=0.35\textwidth]{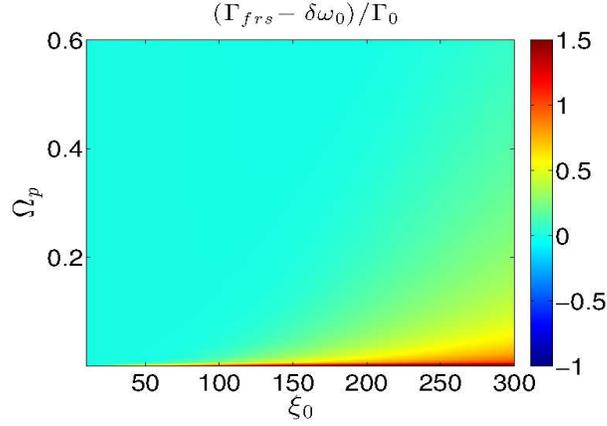}
\caption{(Color online) Growth rate ratio $(\Gamma_{\text{frs}}-\delta\omega_0)/\Gamma_{0}$ of the FRS in the presence $(\Gamma_{\text{frs}}-\delta\omega_{0})$, and in the absence of the RR force $\Gamma_0$ as a function of the normalized plasma density $\Omega_p\equiv \omega_p/\omega_{0r}$ and normalized pump laser amplitude $\xi_0 = e A_0 / m$. The normalized growth rate is plotted on log$_{10}$ scale.}
\label{fig1}
\end{figure}

For the estimation of the FRS growth rate in a low-density plasma, $\omega_{p}^{'} \ll \omega_{0r}$, both the Stokes and the anti-Stokes modes have to be taken into account  \cite{kumar13}. Substituting the pump laser frequency shift $\delta \omega_0$ and ignoring the finite ${k}_{\perp}$ gives $D_{\pm}= (\omega \pm \omega_{0r})^2-\omega_{p}^{'2}-(k_z\pm k_0)^2 $. On expressing $\omega=\omega_p^{'}+i\Gamma_{\text{frs}}$, where $\Gamma_{\text{frs}}$ is the growth rate of the FRS instability, yields $D_{\pm}\approx 2 i \Gamma_{\text{frs}} (\omega_p^{'}\pm \omega_{0r}),\,D_e\approx 2 i \omega_p^{'} \Gamma_{\text{frs}}$.  On assuming $k_z^2 \approx \omega_p^{'2},\, \omega_{p}^{'2} - \omega_{0r}^2\approx -\omega_{0r}^2$, 
we get, in the weakly-coupled regime $\Gamma_{\text{frs}} \ll \omega_p^{'}$, the growth rate of the FRS as
\begin{gather}
\Gamma_{\text{frs}}= -\frac{\omega_{p}^{2}\varepsilon_{\text{rad}} \xi_0^2}{2\omega_{0r}} \pm \frac{\omega_{p}^{2} \xi_0 \text{cos}(\theta/2)}{\sqrt{8}\gamma_0^2\omega_{0r}} \sqrt[4]{(1+2\varepsilon_{\text{rad}}^2 \xi_0^2 \gamma_0^4 )^2 + \left(\frac{\varepsilon_{\text{rad}} \xi_0^2 \gamma_0\omega_{0r}}{\omega_{p}^{'}}\right)^2},\nonumber \\
\text{tan}\theta = \left(\frac{-\varepsilon_{\text{rad}} \xi_0^2 \gamma_0(\omega_{0r}/\omega_{p}^{'})}{(1+2\varepsilon_{\text{rad}}^2 \xi_0^2 \gamma_0^4 )}\right).
\label{frs_growth}
\end{gather} 
\noindent
Without the RR force $\varepsilon_{\text{rad}}=0$,  one recovers the relativistic growth rate of the FRS instability as derived before \cite{decker:2047,Gibbon:2005ys}. Fig.\ref{fig1} shows the growth rate ratio of the FRS with $(\Gamma_{\text{frs}}-\delta\omega_0)$ and without $(\Gamma_0)$ RR force. It is evident that the RR force strongly enhances the growth rate of the FRS at lower plasma densities $\omega_p^{'} / \omega_{0r} \ll 1$ and higher laser amplitude $\xi_0 \gg 1$, which is also apparent from Eq.\eqref{frs_growth}. The strong growth enhancement due to the RR force is counterintuitive as the later is generally considered as a damping force similar to collisions in plasmas. One can attribute this enhancement in the growth rate of the FRS due to the mixing between the Stokes and the anti-Stokes modes mediated by the RR force. Without the RR force, nonlinear currents driving the Stokes and the anti-Stokes 
modes have opposite polarizations. Since the phase shift induced by the RR force is polarization dependent, it is opposite for the Stokes and the anti-Stokes modes. The opposite phase shifts, consequently, lead to the interaction between the nonlinear current terms and phase shifts accumulation in Eq.\eqref{disp_rel}. This phase shift accumulation is termed  as the manifestation of the nonlinear mixing of the two modes, and it is responsible for the enhanced growth rate of the FRS instability. Intuitively this growth enhancement can be imagined to occur due to the availability of an additional channel of the  laser 
energy decay due to the RR force induced damping and its subsequent utilization by both the Stokes and the anti-Stokes modes.

Since, the resonant excitation of both the Stokes and the anti-Stokes modes is the essential condition for the growth enhancement of the FRS, let us estimate the conditions under which both the modes are excited. Resonant excitation of the Stokes modes $(D_{-}=0)$ is always possible due to kinematical considerations.  However, the simultaneous resonant excitation of both the modes is only possible in a tenuous plasma $(\omega_p^{'}\ll \omega_{0r})$. The resonant excitation of the Stokes mode leads to frequency mismatch for the anti-Stokes mode defined as $\Delta \omega_m = \omega_p^{'}+\omega_{0r} -[\omega_p^{'2}+(k_p^{'}+k_0)^2 + D_{+}]^{1/2}$, and it reads as $\Delta\omega_m=-\omega_p^{'3}/\omega_{0r}^2+9 \omega_p^{'4}/4 \omega_{0r}^3$. As shown in Ref. \cite{kumar13}, this frequency mismatch is smaller than the actual growth rate $\Gamma_{\text{frs}}-\delta\omega_0$ of the FRS 
instability. This makes the inclusion of both the modes important while deriving the growth rate of the FRS. The RR force only marginally enhances the growth rate of the FRS, if only the Stokes mode is resonantly excited in the plasma. This can be understood easily as the nonlinear  mixing of the two Raman sidebands is absent in this case. The phase shift caused by the RR force maintains the laser energy transfer to the Stokes mode for a longer time causing minor enhancement in the growth rate of the FRS.

These results are important for the ultra-intense laser-plasma interaction as the 
onset of parametric instabilities  appears again, changing the frequency spectra and shapes of extremely intense short laser pulses.  Moreover, enhanced FRS of the laser pulse provides an alternative way to detect the RR effects on the spectra of low-energy optical photons. This is in contrast to the scheme of the nonlinear Compton scattering of a counter-propagating relativistic electron in a strong laser field, which aims to discern the signatures of the RR force on the spectra of high-energy gamma-ray photons \cite {Di_Piazza_2012}.

\section{Conclusions}

To summarize, we have investigated RR effects for the collision of an ultra-intense laser pulse with ultra-relativistic electrons in the quantum regime as well as with a plasma in the classical regime. In contrast to the classically predicted narrowing of the energy width of particle beams, we have shown by employing a kinetic approach that the stochastic nature of photon emission spreads up the energy distribution of the electrons, if quantum effects are substantial. Further, the quantum computation including two-photon emission indicates an extensive broadening of the predicted angular range of the emitted radiation. In the quantum regime this can be associated with the discontinuous energy loss of the electron and the following modification of its trajectory, whereas in classical electrodynamics RR effects alter an electron's trajectory smoothly. Moreover, the forward Raman scattering in plasmas is shown to be significantly amplified by the inclusion of classical RR, due to the 
induced nonlinear mixing of the anti-Stokes and the Stokes modes. Finally, our numerical examples indicate that all the discussed effects should be detectable with laser-based electron accelerators with presently available and next-generation lasers.


\vspace{.5cm}

\begin{thebibliography}{unsrt}
\bibliographystyle{nature}
\bibitem{Di_Piazza_2012} Di Piazza A, M\"uller C, Hatsagortsyan K Z and Keitel C H 2012 Rev. Mod. Phys. {\bf 84} 1177 
\bibitem{Rohrlich:2002} Rohrlich F 2002 Phys. Lett. A \textbf{303} 307
\bibitem{Bonifacio_1984} Bonifacio R and Casagrande F 1984 Opt. Comm. \textbf{50} 251  
\bibitem{Bonifacio_1985} Bonifacio R and Casagrande F 1985 Nucl. Instr. Meth. Phys. Res. A \textbf{237} 168 
\bibitem{Moshammer_2009} Moshammer R and Ullrich J 2009 J. Phys. B: At. Mol. Opt. Phys. \textbf{42} 130201 
\bibitem{Apyan_2005} Apyan A et al. 2005 Nucl. Instr. and Meth. B, \textbf{234} 128
\bibitem{Landau_b_2_1975} Landau L D and Lifshitz E M 1975 \textit{The classical theory of fields} (Oxford: Butterworth-Heinemann)
\bibitem{Spohn:2000}  Spohn H 2000 Europhys. Lett. \textbf{50} 287 
\bibitem{Spohn:2004}  Spohn H 2004 \textit{Dynamics of Charged Particles and Their Radiation Field} (Cambridge: Cambridge University Press)
\bibitem{Rohrlich_b_2007} Rohrlich F 2007 \textit{Classical Charged Particles} (Singapore: World Scientific)
\bibitem{Bulanov2011a} Bulanov S V, Esirkepov T Zh, Kando M, Koga J K and Bulanov S S 2011 Phys. Rev. E {\bf 84} 056605  
\bibitem{DiPiazza:2008} Di Piazza A 2008 Lett.\ Math.\ Phys.\ {\bf 83} 305 
\bibitem{Tamburini_2013} Tamburini M, Keitel C H and Di Piazza A 2013 Electron dynamics controlled via self-interaction \textit{Preprint} arXiv:1306.3328
\bibitem{Di_Piazza_2009} Di Piazza A, Hatsagortsyan K Z and Keitel C H 2009 Phys. Rev. Lett. {\bf 102} 254802 
\bibitem{Heinzl2013} Heinzl T, Harvey C, Ilderton A, Marklund M, Bulanov S S, Rykovanov S, Schroeder C B, Esarey E and Leemans W P 2013 Detecting radiation reaction at moderate laser intensities \textit{Preprint} arXiv:1310.0352
\bibitem{Bulanov2011}  Bulanov S V et al. 2011 Nucl. Instr. Meth. Phys. Res. A \textbf{660} 31 
\bibitem{Zhidkov_2013} Zhidkov A, Masuda S, Bulanov S S, Hosokai T, Koga J and R. Kodama 2013 Radiation Reaction Effects in Cascade Scattering of Intense, Tightly Focused Laser Pulses by Relativistic Electrons \textit{Preprint} arXiv:1308.1608
\bibitem{Ilderton2013} Ilderton A and Torgrimsson G 2013 Radiation reaction from QED: lightfront perturbation theory in a plane wave background \textit{Preprint} arXiv:1304.6842
\bibitem{Boca2009} Boca M and Florescu V 2009 Phys. Rev. A \textbf{80} 053403
\bibitem{Mackenroth2011} Mackenroth F and Di Piazza A 2011 Phys. Rev. A \textbf{83} 032106 
\bibitem{Seipt2011} Seipt D and K\"ampfer B 2011 Phys. Rev. A \textbf{83} 022101
\bibitem{Krajewska2012} Krajewska K and Kami\'nski J Z 2012 Phys. Rev. A \textbf{85} 062102
\bibitem{Harvey_2012} Harvey C, Heinzl T, Ilderton A and Marklund M 2012 Phys. Rev. Lett. \textbf{109} 100402
\bibitem{Jackson_b_1975} Jackson J D 1975 \textit{Classical Electrodynamics} (New York: Wiley)
\bibitem{DiPiazza:2010mv} Di Piazza A, Hatsagortsyan K Z and Keitel C H 2010  Phys.\ Rev.\ Lett.\  {\bf 105} 220403 
\bibitem{Lotstedt2009a} L\"otstedt E and Jentschura U D 2009 Phys. Rev. A \textbf{80} 053419
\bibitem{Lotstedt2009b} L\"otstedt E and Jentschura U D 2009 Phys. Rev. Lett. \textbf{103} 110404
\bibitem{Seipt2012} Seipt D and K\"ampfer B 2012 Phys. Rev. D \textbf{85} 101701
\bibitem{Mackenroth2013} Mackenroth F and Di Piazza A 2013 Phys. Rev. Lett \textbf{110} 070402 
\bibitem{kumar13} Kumar N, Hatsagortsyan K Z and Keitel C H 2013 Phys. Rev. Lett. {\bf 111} 105001 
\bibitem{Zhidkov_2002} Zhidkov A, Koga J, Sasaki A and Uesaka M 2002 Phys. Rev. Lett. \textbf{88} 185002
\bibitem{Naumova_2009} Naumova N M, Schlegel T, Tikhonchuk V T, Labaune C, Sokolov I V and Mourou G A 2009 Phys. Rev. Lett. \textbf{102} 025002 
\bibitem{Chen_2010} Chen M, Pukhov A, Yu T P and Sheng Z M 2011 Plasma Phys. Controlled Fusion {\bf 53} 014004 
\bibitem{Tamburini_2010}  Tamburini M, Pegoraro F, Di Piazza A, Keitel C H and Macchi A 2010 New J. Phys. {\bf 12} 123005
\bibitem{Tamburini_2011}  Tamburini M, Pegoraro F, Di Piazza A, Keitel C H, Liseykina T V and Macchi A 2011, Nucl. Instr. Meth. Phys. Res. {\bf 653} 181 
\bibitem{Baier_b_1998} Baier V N, Katkov V M and Strakhovenko V M 1998, \textit{Electromagnetic processes at high energies in oriented single crystals} (Singapore: World Scientific)
\bibitem{Khokonov_2004} Khokonov M Kh 2004 Soviet Phys. JETP \textbf{99} 690 
\bibitem{Sokolov:2010am} Sokolov I V, Naumova N M, Nees J A and Mourou G A 2010 Phys.\ Rev.\ Lett.\  {\bf 105} 195005 
\bibitem{Yanovsky_2008} Yanovsky V \textit{et al.} 2008 Opt. Express {\bf 16} 2109
\bibitem{Ritus1985} Ritus V I 1985 J. Sov. Laser Res. \textbf{6} 497
\bibitem{Gardiner_b_2009} Gardiner C 2009 \textit{Stochastic Methods: A Handbook for the Natural and Social Sciences}  (Berlin: Springer) 
\bibitem{Lifshitz_1981} Lifshitz E M and Pitaevskii L P 1981 \textit{Physical Kinetics} (Oxford: Pergamon Press)
\bibitem{Neitz2013} Neitz N and Di Piazza A 2013 Phys. Rev. Lett. \textbf{111} 054802 
\bibitem{ERL} Bazarov I V, Bilderback D H, Gruner S M, Padamsee H S, Talman R, Tigner M, Krafft G A, Merminga L and Sinclair C K 2001 \textit{Proc. Particle Accelerator Conference 2001} 230 
\bibitem{Leemans_2006} Leemans W P, Nagler B, Gonsalves A J, T\'oth Cs, Nakamura K, Geddes C G R, Esarey E, Schroeder C B and Hooker S M 2006 Nature Phys. \textbf{2} 696 
\bibitem{Clayton_2010} Clayton C E et al. 2010 Phys. Rev. Lett. \textbf{105} 105003 
\bibitem{Furry_1951} Furry W H 1951 Phys. Rev. \textbf{81} 115 
\bibitem{Fradkin_Gitman} Fradkin E S, Gitman D M and Shvartsman S M 1991 \textit{Quantum Electrodynamics} (Berlin: Springer)
\bibitem{Landau_b_4_1982} Berstetskii V B, Lifshitz E M and Pitaevskii L P 1982 \textit{Quantum Electrodynamics} (Oxford: Elsevier)
\bibitem{Mackenroth2010} Mackenroth F, Di Piazza A and Keitel C H 2010 Phys. Rev. Lett. \textbf{105} 063903 
\bibitem{Kruer:2003yq} Kruer W 2003, \textit{The Physics of Laser Plasma Interactions}, Frontiers in Physics (Boulder, CO: Westview)
\bibitem{Brueckner:1974uq} Brueckner K A and Jorna S 1974 Rev. Mod. Phys. {\bf 46} 325 
\bibitem{drake:778} Drake J F, Kaw P K, Lee Y C, Schmid G, Liu C S and Rosenbluth M N 1974 Phys. Fluids {\bf 17} 778 
\bibitem{tripathi:468} Tripathi V K and Liu C S 1991 Phys. Fluids. B {\bf 3} 468 
\bibitem{mckinstrie:2626} McKinstrie C J and Bingham R 1992 Phys. Fluids. B {\bf 4} 2626 
\bibitem{Sakharov:1994fk} Sakharov A S and Kirsanov V I 1994 Phys. Rev. E {\bf 49} 3274
\bibitem{decker:2047} Decker C D, Mori W B, Tzeng K C and Katsouleas T 1996 Phys. Plasmas {\bf 3} 2047 
\bibitem{Barr:1999ve} Barr H C, Mason P and Parr D M 1999 Phys. Rev. Lett. {\bf 83} 1606 
\bibitem{jr.:1440}  Antonsen J T M and Mora P 1993 Phys. Fluids. B {\bf 5} 1440 
\bibitem{guerin:2807} Guerin S, Laval G, Mora P, Adam J C, Heron A and Bendib A 1995 Phys. Plasmas {\bf 2} 2807
\bibitem{Quesnel:1997fk} Quensel B, Mora P, Adam J C, Guerin S, Heron A and Laval G 1997 Phys. Rev. Lett. {\bf 78} 2132 
\bibitem{Esarey:2009fk} Esarey E, Schroeder C B and Leemans W P 2009 Rev. Mod. Phys. {\bf 81} 1229 and references therein
\bibitem{Schlegel:2012bh} Schlegel T and Tikhonchuk V T 2012 New. J. Phys. {\bf 14} 073034 
\bibitem{Sokolov:2010kl} Sokolov I V, Nees J A, Yanovsky V P, Naumova N M and Mourou G A 2010 Phys. Rev. E {\bf 81} 036412
\bibitem{Keitel:1998fk} Keitel C H, Szymanowski C, Knight P L and Maquet A 1998 J. Phys. B {\bf 31} L75 
\bibitem{eli:kx} http://www.extreme-light-infrastructure.eu
\bibitem{Gibbon:2005ys} Gibbon P 2005, \textit{Short Pulse Laser Matter Interaction with Matter:An Introduction} (Singapore: World Scientific)
\bibitem{Akhiezer:1956mz} Akhiezer A I and Polovin R 1956 Sov. Phys. JETP {\bf 3} 696 
\bibitem{S.V.-Bulanov:2001kx} Bulanov S V \textit{et al.} 2001 \textit{Reviews of Modern Physics} Vol.22 edited by Shafranov V D (New York: Kluwer/Plenum) 
 
\end{thebibliography}
\end{document}